\documentclass[preprint]{aastex}

\usepackage{ifthen}
\usepackage{natbib}
\input{psfig.sty}

\def\km{{\rm\,km}}
\def\kms{\km ${\rm\, s}^{-1}$}

\def\pmpc{{\rm\,Mpc$^{-1}$}}
\def\msun{{\rm\,M_\odot}}

\newcommand{\Line}[0]{%
  \rule{0cm}{0cm}\\\hrule\rule{0cm}{0cm}%
}

\begin{document}

\Line

\noindent {\bf \large A class of compact dwarf galaxies
from disruptive processes in galaxy clusters.}

\begin{flushleft}

{M. J. Drinkwater$^*$,
M. D. Gregg$^{\dagger}$,
M. Hilker$^{\ddagger}$,
K. Bekki$^\S$
W. J. Couch$^\S$
H. C. Ferguson$^{\parallel}$
J. B. Jones$^{\P}$
\& S. Phillipps$^{\#}$  }

\bigskip
\noindent Published: 2003, Nature, 423, 519-521
\bigskip
{

{$^*$
Department of Physics, 
University of Queensland, 
Queensland 4072,Australia}\\
{$^{\dagger}$ Department of Physics, University of California, Davis,
CA95616 and Institute for Geophysics and Planetary Physics, Lawrence
Livermore National Laboratory, L-413, Livermore, CA 94550, USA}\\
{$^{\ddagger}$ Sternwarte der Universit\"at Bonn, Auf dem H\"ugel 71,
53121 Bonn, Germany}\\
{$^\S$ School of Physics, University of New South Wales, Sydney 2052,
Australia}\\
{$^{\parallel}$ Space Telescope Science Institute, 3700 San Martin
Drive, Baltimore, MD 21218, U.S.A.}\\
{$^{\P}$ School of Physics and Astronomy, University of Nottingham,
University Park, Nottingham NG7 2RD, U.K.}\\
{$^\#$ Astrophysics Group, Department of Physics, University of Bristol,
Tyndall Avenue, Bristol BS8 1TL, U.K.}

}

\end{flushleft}

\vskip -1.25cm
\Line
\vskip 0.2cm



{\bf Dwarf galaxies have attracted increased attention in recent
years, because of their susceptibility to galaxy transformation
processes within rich galaxy clusters\citep{moo1996,bas1994,bek2001}.
Direct evidence for these processes, however, has been difficult to
obtain, with a small number of diffuse light trails\citep{gre1998} and
intra-cluster stars\citep{dur2002,for2002} being the only signs of
galaxy disruption. Furthermore, our current knowledge of dwarf galaxy
populations may be very incomplete, since traditional galaxy surveys
are insensitive to extremely diffuse or compact
galaxies\citep{dis1976}. Aware of these concerns, we undertook a novel
all-object survey of the Fornax galaxy cluster\citep{dri2000}. This
revealed a new population of compact members\citep{dri2000a,phil2001},
completely overlooked in previous conventional surveys. Here we
demonstrate that these ``ultra-compact'' dwarf galaxies are
structurally and dynamically distinct from both globular star clusters
and known types of dwarf galaxy, and thus represent a new class of
dwarf galaxy.  Our data are consistent with the interpretation that
these are the remnant nuclei of disrupted dwarf galaxies, making them an
easily observed tracer of galaxy disruption.}

We used the Two Degree Field (2dF) multi-object spectrograph on the
3.9m Anglo-Australian Telescope to make the first ever large {\em
all-object} spectroscopic survey of the Fornax galaxy
cluster\citep{dri2000}. The 400-fibre multiplex advantage of the 2dF
allowed us to include unresolved objects (normally ignored as
``stars''). Among these ``stars'' we
discovered\citep{dri2000a,phil2001} a population of objects in the
cluster with luminosities of {$M_V \approx -11$ mag}, intermediate
between typical globular star clusters ($M_V \approx -8$ mag) and
normal dwarf galaxies ($M_V \approx$ $-$11 to $-$16 mag).  These
``ultra-compact'' dwarf (UCD) galaxies are unlike other dwarf galaxies
of similar luminosity by virtue of their high central concentration
and thus star-like morphology in typical 1 arcsecond resolution
ground-based imaging.  At the distance of the Fornax
Cluster\citep{dri2001b} (20 Mpc; Hubble constant $H_0=75$\kms \pmpc)
this implies sizes of, at most, 100 pc. We proposed\citep{dri2000a}
that the UCDs were either unusually large and isolated star clusters
or a new type of compact galaxy perhaps formed by the disruption of
nucleated dwarf galaxies.  The UCDs are more luminous than Galactic
globular star clusters, but they could be the extreme bright tail of
the populous globular cluster system around the central galaxy of the
Fornax Cluster NGC~1399\citep{mie2002}.  The UCD luminosities,
however, also suggestively overlap the luminosity distribution of the
nuclei of dwarf elliptical galaxies (which peaks at $M_V\approx -10$
mag\citep{lot2001}).  A third possibility is that UCDs are simply
ordinary nucleated dwarf ellipticals, but at one extreme of a
continuous sequence of envelope luminosities, having envelopes too
faint to be easily detected.  The lack of extended halos around
the UCDs in photographic images\citep{phil2001} argued against this
third hypothesis.  In this letter we present new observations of the
UCDs aimed at determining their nature.  Our deep imaging of the
Fornax Cluster confirms that UCDs are morphologically distinct from
ordinary dwarf ellipticals while our high-resolution imaging and
spectroscopy show that they are not globular star clusters, so they
represent a new class of dwarf galaxy. Our interpretation is that the
UCDs are the stripped nuclei of dwarf elliptical galaxies and as such
are a new tracer of galaxy disruption processes in clusters.

That normal dwarf galaxies and UCDs are quite distinct,
morphologically, is shown in Fig.~1 where we present an envelope
surface brightness versus core luminosity plot. Nucleated dwarf
elliptical (dE,N) galaxies show a correlation between the effective
surface brightness of their envelopes and their core luminosities. In
contrast, UCDs lie in a completely different region of this diagram:
since they have no detectable outer envelope, only upper limits can be
put on their surface brightness, all of which lie 3.5 magnitudes (a
factor of 25) fainter than the dE,N galaxies with the same nuclear
luminosity. Any intermediate objects would have easily been detected
in our survey so UCDs are clearly not part of the normal distribution
of dwarf elliptical galaxies.  The UCDs are also more concentrated to
the centre of the cluster than the dwarf galaxies\citep{dri2000a}; the
radial distribution of the 7 known UCDs differs from that of the 37
dE,N galaxies in our survey at the 99.7\% confidence level. We
therefore conclude that UCDs are not extreme examples of normal dwarf
galaxies with very faint envelopes.

We obtained high-resolution imaging with the {\it Hubble Space
Telescope} (HST) of the first five UCDs discovered.  We also observed
a normal nucleated dwarf elliptical (dE,N) galaxy for
comparison. Further comparisons are provided by similar published
observations of six dwarf elliptical galaxies in the Virgo
Cluster\citep{geh2002}.  Our images are shown in Fig.~2. The UCDs are
very compact, typically only a few times larger than the 0.05 arc
second image resolution.  The UCD profiles were well-fitted by both
King models (normal for globular star clusters) and de Vaucouleurs
R$^{1/4}$ law profiles (normal for giant elliptical galaxies) but are
not consistent with the exponential profiles typical of normal dwarf
galaxies (excluded at the 99.9\%\ confidence level). The effective
radii (defined to contain half the light) of the UCDs range from 10 to
22 parsecs. In the case of the most luminous UCD, UCD3, the model
profile required an additional, but still very small (60 pc scale
length), exponential halo component for a good fit. As expected, the
comparison nucleated dwarf, FCC~303, also required a two-component
model of a core plus a larger (300 pc scale length) exponential halo.
The UCD cores are all significantly larger than Galactic globular star
clusters (typical effective radii of 3--5 pc) and are also larger than
the 8 pc core of the nucleated dwarf, FCC~303. Conversely, the UCDs
are all smaller than known dwarf galaxies. Even the halo we do detect
around UCD3 is tiny for a dwarf elliptical; the most compact normal
dwarf galaxies in the Virgo Cluster have scale lengths of 160 pc
\citep{dri1991}.

To study the internal dynamics and estimate the masses of the UCDs and
FCC~303, we measured their internal velocity dispersions with the Very
Large Telescope (VLT) and the W.M. Keck-II Telescope. The velocity
dispersions range from 24 to 37 \kms, considerably higher than those
of Galactic globular star clusters, but overlapping the values published
for the most luminous globular star clusters in the spiral galaxy M~31
\citep{djor1997}.  Assuming the UCDs are dynamically relaxed systems,
we can use their sizes and velocity dispersions to estimate their
masses. Using a King model mass estimator\citep{mey2001} for our
profile fits we obtain masses of 1--5$\times10^7\msun$ (Solar masses)
for the UCDs and 1.4$\times10^7\msun$ for the dE,N galaxy nucleus,
compared to around $4\times10^6\msun$ for the most massive globular
star clusters\citep{djor1997}. We obtained similar masses using
Plummer models\citep{her1987}, although these profiles did not fit our
data as well.  Using the HST data to measure the total V-band
luminosities of the UCDs, we derived their mass-to-light ratios.  The
mass-to-light ($M/L$) ratios range from 2 to 4 in Solar units,
compared to around unity for typical globular star clusters. The
largest globular star clusters in M~31 have $M/L=$1--2.  In summary,
the UCDs have luminosities, sizes and velocity dispersions similar to
the cores of nucleated dwarf galaxies (in both the Fornax and
Virgo\citep{geh2002} clusters), but are larger and have higher
mass-to-light ratios than even the largest globular star clusters.

The distinction between UCDs and other types of stellar system is made
even clearer if we consider the dynamical correlations between the
physical observables of luminosity and velocity dispersion shown in
Fig.~3.  The figure compares UCDs with globular clusters, nucleated
dwarf elliptical galaxies and giant elliptical galaxies. We also
include the large Milky Way globular cluster, $\omega$ Centauri, which
is thought to have originated outside the Milky Way\citep{gne2002}.
The UCDs lie well off the globular cluster $L \propto \sigma^{1.7}$
relation\citep{djor1997}, filling a previously unoccupied part of the
diagram, but on an extrapolation of the elliptical galaxy $L \propto
\sigma^{4}$ Faber-Jackson law\citep{fab1976}.  The properties of the
cores of the comparison dE,N galaxies are also shown: these are
distributed between the globular clusters and the UCDs.  The locations
of the UCDs and the dwarf galaxy nuclei strongly support the ``galaxy
threshing'' hypothesis: UCDs are nucleated dwarf galaxies whose outer
envelopes have been tidally stripped by interactions with the central
cluster galaxy\citep{bek2001}.  Removing the halo of a normal dE,N
galaxy reduces the total luminosity by a factor of $\approx 100$ (5
mag), but barely changes the central velocity dispersion and nuclear
luminosity (the dynamical relaxation time of the remaining core can be
as long as $10^{11}$ years). The predicted effect of threshing is
shown as a dotted line in Fig.~3 connecting the positions of FCC303
and that of its nucleus alone. The case for tidal stripping is further
supported by our detection of a small halo around the largest
UCD. This object may be partially stripped, caught at an intermediate
stage between UCDs and dE,N galaxies. As threshing will have been at
least as important at earlier times in the life of the cluster, this
implies a much larger population of UCDs than the 7 currently known,
although if they have the same luminosity function as dwarf galaxy
nuclei\citep{lot2001}, most will be fainter than the limit of our
survey.

Our new observations confirm that UCDs are qualitatively different to
both nucleated dwarf galaxies and the most luminous globular star
clusters. Even if UCDs were formed from dE,N galaxies, as proposed by
the threshing hypothesis, they constitute a new and distinct type of
galaxy, just as spirals stripped of their star forming abilities
constitute the galaxy type S0\citep{bek2002}, and merging disk
galaxies are believed to form ellipticals. The discovery of these
objects confirms the prediction\citep{dis1976} that very compact
galaxies exist which were previously misclassified as foreground
stars. Although we have shown that UCDs are dynamically distinct from
globular star clusters, the two can only be distinguished at the
distance of the Fornax Cluster by high resolution imaging and
spectroscopy.  Any UCDs captured by the central elliptical may
therefore be mistaken for ordinary globular star clusters in existing
studies. The two populations could be separated, in principle, by
observations similar to those we described here to place them in
Fig.~3, but the requisite spectroscopy for fainter objects would be
challenging with existing 8--10 m telescopes. Any dismantled dwarf
elliptical galaxies captured by the central galaxy will also
contribute ordinary globular star clusters to the mix.  Together,
these processes may explain the unusually high frequency of globular
star clusters in the central elliptical galaxies of galaxy
clusters\citep{hil1999c} because these dwarf galaxies also contain
higher than average globular cluster populations\citep{mil1998}.
Searches for UCDs in other rich environments, the populous Virgo
Cluster as well as dense groups, will clarify the nature and origin of
UCDs while also potentially exploring the role of tidal disruption
processes in galaxy and cluster evolution.

\addtolength{\baselineskip}{-0.05\baselineskip}
\bibliographystyle{unsrtnat}
\bibliography{mjdrefs}

\begin{thebibliography}{28}
\expandafter\ifx\csname natexlab\endcsname\relax\def\natexlab#1{#1}\fi
\expandafter\ifx\csname url\endcsname\relax
  \def\url#1{{\tt #1}}\fi

\bibitem[{Moore} et~al.(1996){Moore}, {Katz}, {Lake}, {Dressler}, and
  {Oemler}]{moo1996}
B.~{Moore}, N.~{Katz}, G.~{Lake}, A.~{Dressler}, and A.~{Oemler}.
\newblock {Galaxy harassment and the evolution of clusters of galaxies.}
\newblock {\em \nat}, 379:\penalty0 613--616, 1996.

\bibitem[{Bassino} et~al.(1994){Bassino}, {Muzzio}, and {Rabolli}]{bas1994}
L.~P. {Bassino}, J.~C. {Muzzio}, and M.~{Rabolli}.
\newblock {Are globular clusters the nuclei of cannibalized dwarf galaxies?}
\newblock {\em \apj}, 431:\penalty0 634--639, August 1994.

\bibitem[{Bekki} et~al.(2001){Bekki}, {Couch}, and {Drinkwater}]{bek2001}
K.~{Bekki}, W.~J. {Couch}, and M.~J. {Drinkwater}.
\newblock {Galaxy Threshing and the Formation of Ultracompact Dwarf Galaxies}.
\newblock {\em \apjl}, 552:\penalty0 L105--L108, May 2001.

\bibitem[{Gregg} and {West}(1998)]{gre1998}
M.~D. {Gregg} and M.~J. {West}.
\newblock {Galaxy disruption as the origin of intracluster light in the Coma
  cluster of galaxies.}
\newblock {\em \nat}, 396:\penalty0 549--552, 1998.

\bibitem[{Durrell} et~al.(2002){Durrell}, {Ciardullo}, {Feldmeier}, {Jacoby},
  and {Sigurdsson}]{dur2002}
P.~R. {Durrell}, R.~{Ciardullo}, J.~J. {Feldmeier}, G.~H. {Jacoby}, and
  S.~{Sigurdsson}.
\newblock {Intracluster Red Giant Stars in the Virgo Cluster}.
\newblock {\em \apj}, 570:\penalty0 119--131, May 2002.

\bibitem[{Ford} et~al.(2002){Ford}, {Peng}, and {Freeman}]{for2002}
H.~{Ford}, E.~{Peng}, and K.~{Freeman}.
\newblock {Extragalactic Planetary Nebulae}.
\newblock In {\em ASP Conf. Ser. 273: The Dynamics, Structure {\&} History of
  Galaxies: A Workshop in Honour of Professor Ken Freeman}, pages 41--+, 2002.

\bibitem[{Disney}(1976)]{dis1976}
M.~J. {Disney}.
\newblock {Visibility of galaxies}.
\newblock {\em \nat}, 263:\penalty0 573--575, October 1976.

\bibitem[{Drinkwater} et~al.(2000{\natexlab{a}}){Drinkwater}, {Phillipps},
  {Jones}, {Gregg}, {Deady}, {Davies}, {Parker}, {Sadler}, and
  {Smith}]{dri2000}
M.~J. {Drinkwater}, S.~{Phillipps}, J.~B. {Jones}, M.~D. {Gregg}, J.~H.
  {Deady}, J.~I. {Davies}, Q.~A. {Parker}, E.~M. {Sadler}, and R.~M. {Smith}.
\newblock {The Fornax spectroscopic survey. I. Survey strategy and preliminary
  results on the redshift distribution of a complete sample of stars and
  galaxies}.
\newblock {\em \aap}, 355:\penalty0 900--914, March 2000{\natexlab{a}}.

\bibitem[{Drinkwater} et~al.(2000{\natexlab{b}}){Drinkwater}, {Jones}, {Gregg},
  and {Phillipps}]{dri2000a}
M.~J. {Drinkwater}, J.~B. {Jones}, M.~D. {Gregg}, and S.~{Phillipps}.
\newblock {Compact stellar systems in the Fornax Cluster: Super-massive star
  clusters or extremely compact dwarf galaxies?}
\newblock {\em Publications of the Astronomical Society of Australia},
  17:\penalty0 227--233, December 2000{\natexlab{b}}.

\bibitem[{Phillipps} et~al.(2001){Phillipps}, {Drinkwater}, {Gregg}, and
  {Jones}]{phil2001}
S.~{Phillipps}, M.~J. {Drinkwater}, M.~D. {Gregg}, and J.~B. {Jones}.
\newblock {Ultracompact Dwarf Galaxies in the Fornax Cluster}.
\newblock {\em \apj}, 560:\penalty0 201--206, October 2001.

\bibitem[{Drinkwater} et~al.(2001){Drinkwater}, {Gregg}, and
  {Colless}]{dri2001b}
M.~J. {Drinkwater}, M.~D. {Gregg}, and M.~{Colless}.
\newblock {Substructure and Dynamics of the Fornax Cluster}.
\newblock {\em \apjl}, 548:\penalty0 L139--L142, February 2001.

\bibitem[{Mieske} et~al.(2002){Mieske}, {Hilker}, and {Infante}]{mie2002}
S.~{Mieske}, M.~{Hilker}, and L.~{Infante}.
\newblock {Ultra compact objects in the Fornax cluster of galaxies: Globular
  clusters or dwarf galaxies?}
\newblock {\em \aap}, 383:\penalty0 823--837, March 2002.

\bibitem[{Lotz} et~al.(2001){Lotz}, {Telford}, {Ferguson}, {Miller},
  {Stiavelli}, and {Mack}]{lot2001}
J.~M. {Lotz}, R.~{Telford}, H.~C. {Ferguson}, B.~W. {Miller}, M.~{Stiavelli},
  and J.~{Mack}.
\newblock {Dynamical Friction in DE Globular Cluster Systems}.
\newblock {\em \apj}, 552:\penalty0 572--581, May 2001.

\bibitem[{Geha} et~al.(2002){Geha}, {Guhathakurta}, and {van der
  Marel}]{geh2002}
M.~{Geha}, P.~{Guhathakurta}, and R.~P. {van der Marel}.
\newblock {Internal Dynamics, Structure, and Formation of Dwarf Elliptical
  Galaxies. I. A Keck/Hubble Space Telescope Study of Six Virgo Cluster Dwarf
  Galaxies}.
\newblock {\em \aj}, 124:\penalty0 3073--3087, December 2002.

\bibitem[{Drinkwater} and {Hardy}(1991)]{dri1991}
M.~{Drinkwater} and E.~{Hardy}.
\newblock {Extreme blue compact dwarf galaxies in the Virgo Cluster}.
\newblock {\em \aj}, 101:\penalty0 94--101, January 1991.

\bibitem[{Djorgovski} et~al.(1997){Djorgovski}, {Gal}, {McCarthy}, {Cohen}, {de
  Carvalho}, {Meylan}, {Bendinelli}, and {Parmeggiani}]{djor1997}
S.~G. {Djorgovski}, R.~R. {Gal}, J.~K. {McCarthy}, J.~G. {Cohen}, R.~R. {de
  Carvalho}, G.~{Meylan}, O.~{Bendinelli}, and G.~{Parmeggiani}.
\newblock {Dynamical Correlations for Globular Clusters in M31}.
\newblock {\em \apjl}, 474:\penalty0 L19--L22, January 1997.

\bibitem[{Meylan} et~al.(2001){Meylan}, {Sarajedini}, {Jablonka}, {Djorgovski},
  {Bridges}, and {Rich}]{mey2001}
G.~{Meylan}, A.~{Sarajedini}, P.~{Jablonka}, S.~G. {Djorgovski}, T.~{Bridges},
  and R.~M. {Rich}.
\newblock {Mayall II=G1 in M31: Giant Globular Cluster or Core of a Dwarf
  Elliptical Galaxy?}
\newblock {\em \aj}, 122:\penalty0 830--841, August 2001.

\bibitem[{Hernquist}(1987)]{her1987}
L.~{Hernquist}.
\newblock {Performance characteristics of tree codes}.
\newblock {\em \apjs}, 64:\penalty0 715--734, August 1987.

\bibitem[{Gnedin} et~al.(2002){Gnedin}, {Zhao}, {Pringle}, {Fall}, {Livio}, and
  {Meylan}]{gne2002}
O.~Y. {Gnedin}, H.~{Zhao}, J.~E. {Pringle}, S.~M. {Fall}, M.~{Livio}, and
  G.~{Meylan}.
\newblock {The Unique History of the Globular Cluster {$\omega$} Centauri}.
\newblock {\em \apjl}, 568:\penalty0 L23--L26, March 2002.

\bibitem[{Faber} and {Jackson}(1976)]{fab1976}
S.~M. {Faber} and R.~E. {Jackson}.
\newblock {Velocity dispersions and mass-to-light ratios for elliptical
  galaxies}.
\newblock {\em \apj}, 204:\penalty0 668--683, March 1976.

\bibitem[{Bekki} et~al.(2002){Bekki}, {Couch}, and {Shioya}]{bek2002}
K.~{Bekki}, W.~J. {Couch}, and Y.~{Shioya}.
\newblock {Passive Spiral Formation from Halo Gas Starvation: Gradual
  Transformation into S0s}.
\newblock {\em \apj}, 577:\penalty0 651--657, October 2002.

\bibitem[{Hilker} et~al.(1999){Hilker}, {Infante}, and {Richtler}]{hil1999c}
M.~{Hilker}, L.~{Infante}, and T.~{Richtler}.
\newblock {The central region of the Fornax cluster. III. Dwarf galaxies,
  globular clusters, and cD halo - are there interrelations?}
\newblock {\em \aaps}, 138:\penalty0 55--70, July 1999.

\bibitem[{Miller} et~al.(1998){Miller}, {Lotz}, {Ferguson}, {Stiavelli}, and
  {Whitmore}]{mil1998}
B.~W. {Miller}, J.~M. {Lotz}, H.~C. {Ferguson}, M.~{Stiavelli}, and B.~C.
  {Whitmore}.
\newblock {The Specific Globular Cluster Frequencies of Dwarf Elliptical
  Galaxies from the Hubble Space Telescope}.
\newblock {\em \apjl}, 508:\penalty0 L133--L137, December 1998.

\bibitem[{Hilker} et~al.(2003){Hilker}, {Mieske}, and {Infante}]{hil2003}
M.~{Hilker}, S.~{Mieske}, and L.~{Infante}.
\newblock {Faint dwarf spheroidals in the Fornax Cluster: A flat luminosity
  function}.
\newblock {\em \aap}, 1:\penalty0 in press, July 2003.

\bibitem[{Ferguson}(1989)]{fer1989}
H.~C. {Ferguson}.
\newblock {Population studies in groups and clusters of galaxies. II - A
  catalog of galaxies in the central 3.5 deg of the Fornax Cluster}.
\newblock {\em \aj}, 98:\penalty0 367--418, August 1989.

\bibitem[{Larsen}(1999)]{lar1999}
S.~S. {Larsen}.
\newblock {Young massive star clusters in nearby galaxies. II. Software tools,
  data reductions and cluster sizes}.
\newblock {\em \aaps}, 139:\penalty0 393--415, October 1999.

\bibitem[{Merritt} et~al.(1997){Merritt}, {Meylan}, and {Mayor}]{mer1997}
D.~{Merritt}, G.~{Meylan}, and M.~{Mayor}.
\newblock {The stellar dynamics of omega centauri.}
\newblock {\em \aj}, 114:\penalty0 1074--1086, September 1997.

\bibitem[{Burstein} et~al.(1997){Burstein}, {Bender}, {Faber}, and
  {Nolthenius}]{bur1997}
D.~{Burstein}, R.~{Bender}, S.~{Faber}, and R.~{Nolthenius}.
\newblock {Global Relationships Among the Physical Properties of Stellar
  Systems.}
\newblock {\em \aj}, 114:\penalty0 1365--1392, October 1997.

\end{thebibliography}

\noindent
{\bf Acknowledgements.} This paper is based on observations made with
the Hubble Space Telescope, the European Southern Observatory Very
Large Telescope, the W.M. Keck Telescope, and the Las Campanas
Observatory 100-inch du Pont Telescope. This work is supported by
the Australian Research Council and the Australian Nuclear Science and
Technology Organisation. MDG acknowledges support from
the National Science Foundation and from the Space
Telescope Science Institute and NASA which is operated by
AURA, Inc.  Part of this work was
performed under the auspices of the U.S. Department of Energy by
University of California Lawrence Livermore National Laboratory.

\noindent
{\bf Correspondence} and requests for materials should be addressed to
M.J.D. (e-mail: mjd@physics.uq.edu.au).

\begin{table}
\caption{Properties of the UCDs (not listed in published Nature paper).}
\bigskip
\begin{tabular}{lclllll}
Name      &  RA (J2000) Dec.\    &$m_V$  &  $M_V$  &$\mu_V$ limit&$\sigma_v$&$R_e$\\
          &  (h,m,s) (d,m,s)     &(mag)  &   (mag) &(mag asec$^{-2}$)
   &(\kms)&(pc) \\
\hline
UCD1       &3:37:03.30 $-$35:38:04.6&19.35& $-$12.2 &   28.5 &  32&    21 \\ 
UCD2       &3:38:06.33 $-$35:28:58.8&19.27& $-$12.2 &   28.5 &  26&    23 \\ 
UCD3       &3:38:54.10 $-$35:33:33.6&17.81& $-$13.7 &   27.5 &  37&    30 \\ 
UCD4       &3:39:35.95 $-$35:28:24.5&19.60& $-$11.9 &   28.0 &  27&    24 \\ 
UCD5       &3:39:52.58 $-$35:04:24.1&19.17& $-$12.3 &   28.5 &  24&    15 \\
FCC303 core&3:45:14.08 $-$36:56:12.4&19.51& $-$12.0 &   ---  &  31&    10 \\  
FCC303     &3:45:14.08 $-$36:56:12.4&14.5 & $-$17.0 &   ---  &  31&   320 \\ 
\end{tabular}

\bigskip

Note: $m_V$ measured from STIS images assuming $B-V=0.9$ mag.
\end{table}

\newpage
\hfil \psfig{file=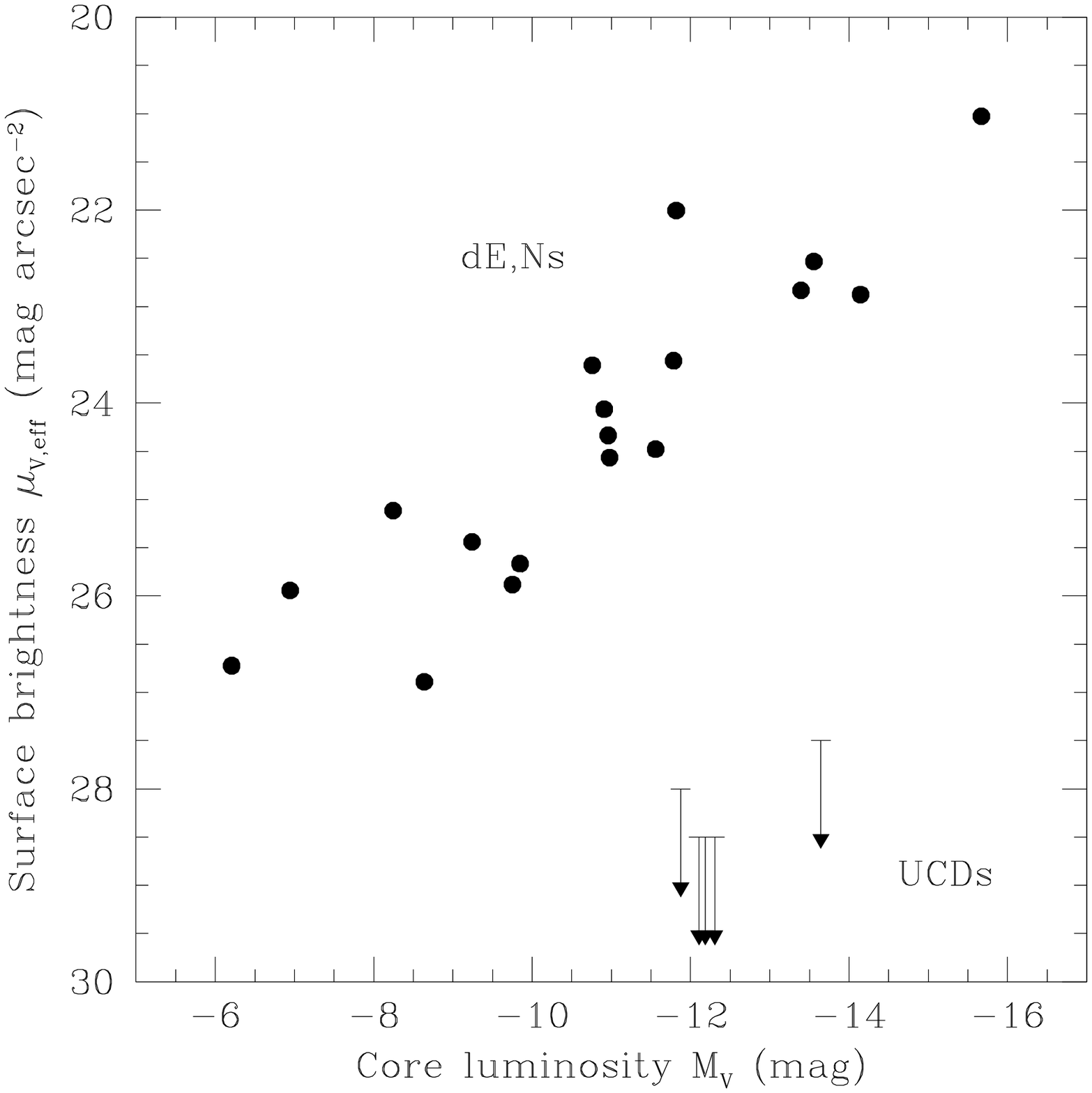,width=8cm}

\noindent

{\bf Figure 1.} Comparison of the morphology of ultra-compact dwarf
(UCD) galaxies with normal nucleated dwarf elliptical (dE,N) galaxies
in the Fornax Cluster. The envelope effective surface brightness in V
of the objects is plotted as a function of the V-band luminosity of
their cores. The luminosity is in units of V-band absolute magnitude
$M_V=const-2.5\log_{10}L$, luminosity $L$ increasing left-to-right.
The surface brightness of the UCDs are upper limits as no envelopes
were detected on scales greater than the image resolution (2
arcseconds full-width half-maximum). The data are from deep images of
2.4 square degrees in the centre of the Fornax Cluster taken with the
100-inch du Pont telescope of the Las Campanas
Observatory\citep{hil2003}. This region contains the original 5 UCDs
as well as 18 nucleated dwarf elliptical galaxies\citep{fer1989}.

\newpage
\hfil \psfig{file=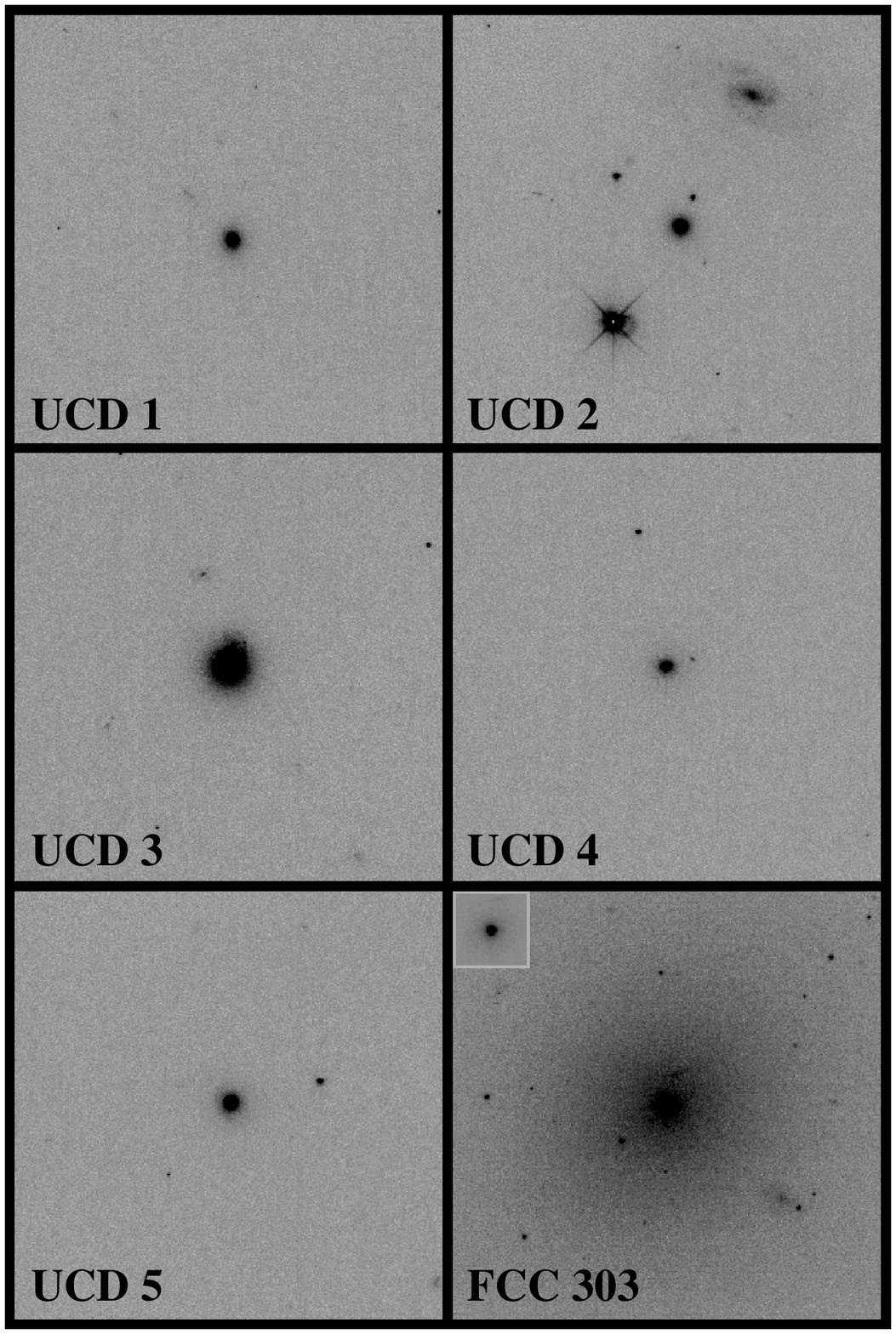,width=8cm}

{\bf Figure 2.} Comparison of {\em Hubble Space Telescope} images of
the ultra-compact dwarf (UCD) galaxies with a normal nucleated dwarf
galaxy (FCC~303\citep{fer1989}).  The panel showing FCC~303 is at the
same contrast as the other five; the inset in the upper left is at a
lower contrast to show the nucleus. The extended halo of FCC~303 more
than fills the 50 arc second field (4.9 kpc or $1.5
\times 10^{20}$m at the Fornax Cluster). These
images, shown in negative format, were obtained in the unfiltered mode
of the Space Telescope Imaging Spectrograph (STIS).  The one-orbit (40
minute) exposures were made in the ``clearpass'' optical CCD imaging
mode of STIS for maximum sensitivity. We measured the UCD sizes with
the ``ishape'' program\citep{lar1999}, which iteratively convolves a
chosen model profile with the STIS point spread function to match the
galaxy profile by minimising a $\chi^2$ difference statistic. We used
a fitting radius of 0.8 arcsec (16 pixels) corresponding to the same
physical radius of 75 pc as used in the Virgo study\citep{geh2002} and
obtained effective radii for the UCDs of 10--22 pc.

\newpage
\hfil \psfig{file=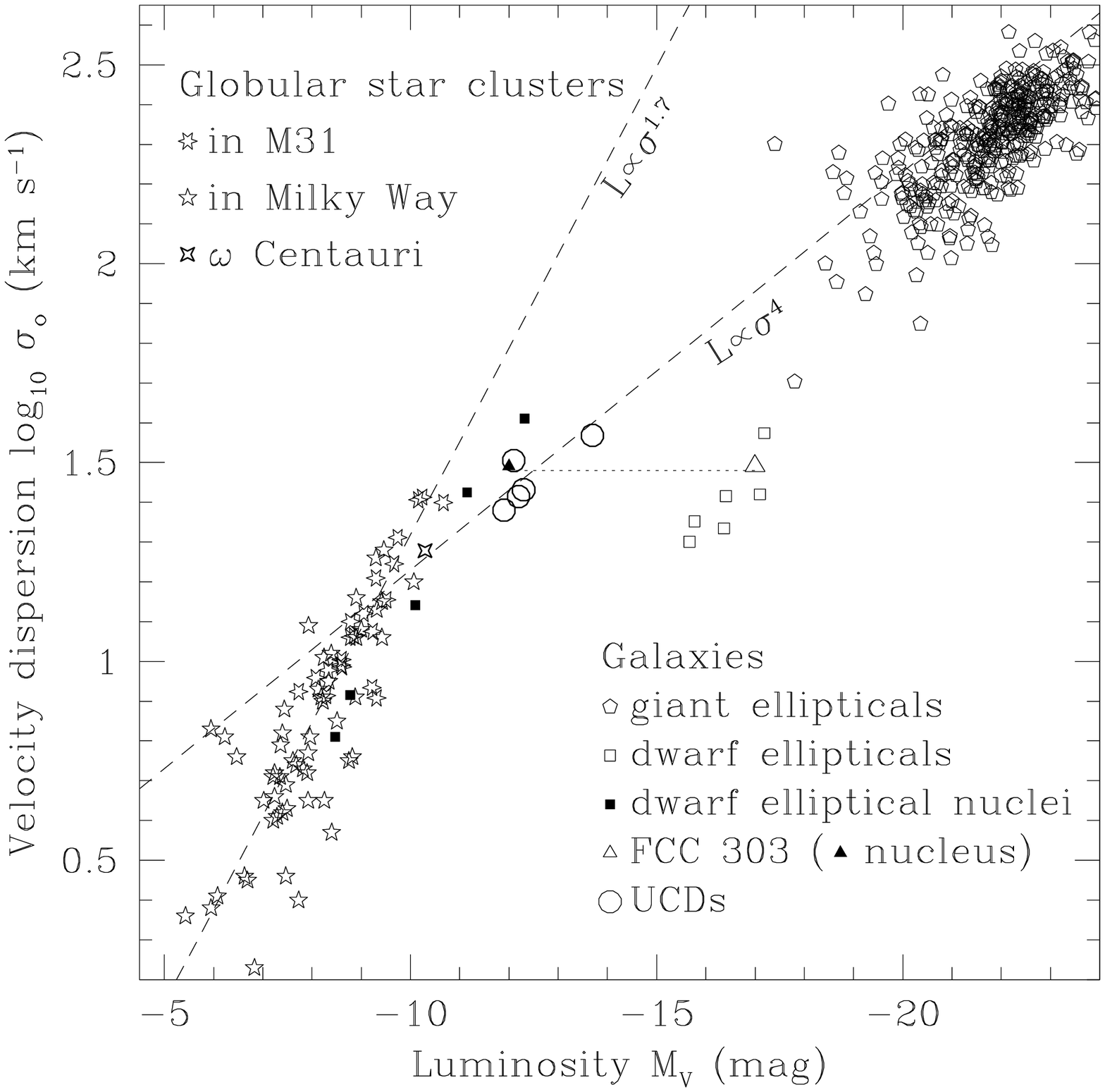,width=8cm}

{\bf Figure 3.} {Comparison of the internal dynamics of the UCDs with
normal galaxies and globular star clusters. The internal velocity
dispersions ($\sigma_0$) of the different objects are plotted as a
function of luminosity (in units of V-band absolute magnitude
$M_V=const-2.5\log_{10}L$, luminosity $L$ increasing
left-to-right). The dashed lines show the Faber-Jackson
relation\citep{fab1976} for elliptical galaxies and the steeper
relation\citep{djor1997} followed by Galactic and M~31 globular
clusters. The dotted lines show the path predicted by our
``threshing'' scenario for nucleated dwarfs being stripped to form
UCDs. The small solid symbols show the locations of dwarf galaxy
nuclei after model subtraction of their parent galaxies.  All the data
are taken from the literature (globular clusters\citep{djor1997},
$\omega$ Centauri\citep{mer1997}, ellipticals\citep{bur1997},
nucleated dwarf ellipticals\citep{geh2002}) except for our
measurements of the UCDs and FCC~303. Note that the UCDs fill a
previously empty region of the diagram between the globular clusters
and galaxies.  The velocity dispersions were obtained from
high-resolution spectra using the Very Large Telescope UV Echelle
Spectrograph and Keck Telescope Echelle Spectrograph and Imager. We
measured velocity dispersions using standard cross-correlations of the
object spectra with stellar template spectra to determine both a
redshift and a correlation width.  The correlation width was used to
estimate the velocity dispersion by comparison with results from
artificially broadening the template stars by convolution with
Gaussian filters of known widths.  }

\end{document}